# Dependable contact related parameter extraction in graphene - metal junctions

Amit Gahoi[a,], Satender Kataria[a], Francesco Driussi[b], Stefano Venica[b], Himadri Pandey[a], David Esseni[b], Luca Selmi[c], M. C. Lemme [a, d*]

[a]RWTH Aachen University, Faculty of Electrical Engineering and Information Technology, Chair of Electronic Devices, Otto-Blumenthal-Str. 2, 52074 Aachen, Germany

[b]Universitá degli Studi di Udine, Dipartimento Politecnico di Ingegneria e Architettura (DPIA),

Via delle Scienze 206, 33100 Udine, Italy

[c]Universitá degli Studi di Modena e Reggio Emilia, Dipartimento di Ingegneria "Enzo Ferrari", Via Vivarelli 10, 41125 Modena, Italy

[d]AMO GmbH, Advanced Microelectronic Center Aachen (AMICA), Otto-Blumenthal-Str. 25, 52074 Aachen, Germany

[*]Corresponding author: max.lemme@eld.rwth-aachen.de, lemme@amo.de

**Abstract** – The accurate extraction and the reliable, repeatable reduction of graphene – metal contact resistance ($R_C$) are still open issues in graphene technology. Here, we demonstrate the importance of following clear protocols when extracting $R_C$ using the transfer length method (TLM). We use the example of back-gated graphene TLM structures with nickel contacts, a complementary metal oxide semiconductor compatible metal. The accurate extraction of $R_C$ is significantly affected by generally observable Dirac voltage shifts with increasing channel lengths in ambient conditions. $R_C$ is generally a function of the carrier density in graphene. Hence, the position of the Fermi level and the gate voltage impact the extraction of $R_C$. Measurements in high vacuum, on the other hand, result in dependable extraction of $R_C$ as a function of gate voltage owing to minimal spread in Dirac voltages. We further assess the accurate measurement and extraction of important parameters like contact-end resistance, transfer length, sheet resistance of graphene under the metal contact and specific contact resistivity as a function of the back-gate voltage. The presented methodology has also been applied to devices with gold and copper contacts, with similar conclusions.

## 1. INTRODUCTION

Graphene exhibits unique and remarkable physical, chemical and electrical properties [1]. It is considered as a promising material for electronic devices and applications. The extraordinary features of graphene have been explored in numerous device demonstrations such as radio frequency (RF) analog transistors [2]–[7], photodetectors [8]–[12], nanoelectromechanical systems [13]–[17] or terahertz modulators [18]–[20]. Low resistivity electrical contacts are fundamental for all of these applications as they provide means for communication between the active devices and the outside world [21]. Because of the low density of states in graphene [22], the charge injection into graphene leads to a high contact resistance ($R_C$) at graphene – metal (G-M) junctions [23], that may be a serious limiting factor for graphene devices. To achieve low contact resistivity, nanostructuring or engineering of graphene under the contact metals have displayed some potential [24]–[26]. Nevertheless, reported values of $R_C$ found in literature vary considerably [27]–[30], which can be attributed to intrinsic factors such as the quality of graphene layers, work functions of the metals ($\phi_M$) or doping of graphene and to extrinsic factors such as specific fabrication procedures. However, there is also a debate on the applicability of standard measurement and extraction methods to obtain dependable and correct $R_C$ values. In particular, the transfer length method (TLM) is widely used to extract $R_C$ from test structures. It has been suggested and demonstrated that the classic TLM may be inappropriate under certain conditions for characterizing G-M contacts [31]–[35]. Since the sheet resistance of the graphene channel ($R_{SH}$) is a dominating component of the total resistance of the device ($R_T$) and since the TLM technique is based on the comparison of the $R_T$ values of the GFETs inside the TLM structure, slight changes in $R_{SH}$ along the TLM array will lead to erroneous $R_C$ values. In addition, the $R_{SH}$ is different from the sheet resistance of graphene under the metal contact ($R_{SK}$) due to the G-M interactions that dope graphene [27], [36], [37] and this should be accounted for in the extraction of the parameters describing the G-M junction.

Here, we present a comprehensive analysis of the TLM technique for extracting $R_C$ in two-dimensional (2D) graphene devices using nickel (Ni) as contact metal. The analysis is further implemented and verified using copper (Cu) and gold (Au) contacts as well for generalization. In the following, we use an extended terminology, where the extracted $R_C$ is called contact front resistance ($R_{CF}$). $R_{CF}$ is extracted from a measured voltage drop at the source/drain contacts that depends on the contact shape and size (i.e. geometry dependent), the charge carrier density ($n_0$) in graphene under the metal contact and the specific contact resistivity $\rho_c$, which is geometrically independent. These additional parameters can be obtained by measuring the contact end resistance ($R_{CE}$) [38], which allows extracting $\rho_c$, the $R_{SK}$ and the transfer length ($L_{TK}$). Furthermore, $R_{SK}$ can be compared to $R_{SH}$ to gain insight into the impact of the contact metal proximity to the graphene on the graphene charge. Finally, the developed method enables the identification of extrinsic effects which significantly affect the extraction of $R_{CF}$ by comparing measurements performed in ambient and vacuum conditions (~ $10^{-6}$ mbar). In particular, the shift of the Dirac voltage ($V_{DIRAC}$) in devices with different channel length ($L_{CH}$) influences the variability of the extracted data.

## 2. THEORY

In the TLM measurement technique, $R_{CF}$ is extracted from the total measured resistance ($R_T$) of the device, which comprises of the sheet resistance of graphene channel ($R_{SH}$), two times $R_{CF}$ and the parasitic resistances (which are generally neglected) of the metal contact pad, probe needles and leads. Here, $R_T$ of a GFET is extracted by applying a voltage $V_{12}$ between contacts 1 and 2 and measuring the current $I_{12}$ as shown in Fig. 1 (a). Therefore,

$$R_T = \frac{V_{12}}{I_{12}} \qquad (1)$$

As mentioned above, the $R_T$ of the device can be deconvoluted into

$$R_T = 2R_{CF} + \frac{R_{SH} * L_{CH}}{W_{CH}} + 2R_M \qquad (2)$$

where $R_M$ is the resistance of the metal that is generally neglected ($R_M \ll R_{SH}$). $R_{CF}$ and $R_{SH}$ are extracted by the TLM extrapolation technique. More details about the TLM extraction technique are reported in the supplementary information (section B). Fig. 1(b) explicitly explains about various resistive components at G-M junction. In Fig. 1(b), for the contact 2, X = 0 corresponds to the front edge of the contact, while X = d is at the end of the contact (where d = 4 μm is the length of the fabricated contact). $R_{CF}$ is defined as the ratio of the voltage drop across the interfacial layer at the front edge of the contact, where the current density is greatest (X=0) as shown in Fig. 1 (b). $R_{CE}$ instead is defined through the voltage drop across the interfacial layer at the edge of the contact, where the current density is least (X=d) as depicted in Fig. 1 (b).

As metal can dope graphene at the contact regions [39], $R_{SK}$ may be different from the $R_{SH}$ [27]. If $R_{SK} \neq R_{SH}$, the extraction of $L_{TK}$, described in detail in the supplementary information (section C), by using the conventional extraction through the TLM method leads to erroneous $L_{TK}$ values. In such a situation, additional measurements are required to extract reliable contact related parameters at the G-M junction, namely the $R_{CE}$ technique [33], [38]. $R_{CE}$ can be measured directly by forcing a known current between contacts 1 and 2 and measuring the generated voltage drop across the G-M stack, while imposing a null current on contact 3 as shown in Fig. 1 (a). In this way, $R_{CE}$ is measured by the use of the additional contact 3, since by forcing a null current, there is no additional voltage drop between contacts 2 and 3. So we have:

$$R_{CE} = \frac{V_{32}}{I_{12}} \qquad (3)$$

The $R_{CE}$ measurement does not depend on the graphene quality outside the metal, which removes errors caused by inhomogeneities of the graphene channel.

Figure 1 (c) is the distributed resistance network describing the transmission line model typically used to describe the G-M contacts. Here, the current is highest at front edge of the contact (X=0) and drops exponentially with the distance (represented by a yellow line as Jv(x)). The "1/e" distance of the voltage drop profile from the front edge of the contact is defined as $L_{TK}$. In layman terms, $L_{TK}$ is the effective electrical length of the contact. The transmission line model equations [38] can be used to extract $\rho_c$, $R_{SK}$ and $L_{TK}$ by using the $R_{CF}$ and $R_{CE}$ values measured with the procedure explained above. Further details on this point are provided in section C of the supporting information.

## 3. RESULTS AND DISCUSSION

Large - area monolayer graphene was grown on a copper (Cu) foil via chemical vapor deposition (CVD) process in a NanoCVD (Moorfield, UK) rapid thermal processing tool [40]. Silicon wafers (p-doped) with resistivity (1-20 Ωμm) were used as a starting substrate and thermally oxidized to achieve a thickness of 85 nm. CVD graphene was transferred onto the silicon/silicon dioxide (Si/SiO$_2$) substrate using an electrochemical delamination technique [41]. Photolithography and reactive ion etching (RIE) were used to define graphene channels using an oxygen plasma process. Nickel (Ni), Copper (Cu) and gold (Au) metals were thermally evaporated to contact graphene, and a lift-off process was carried out in order to define source-drain contact pads. Subsequently, rapid thermal annealing (RTA) process was carried out for 2 hours in an argon (95%)/hydrogen (5%) atmosphere at 450°C to minimize the PMMA residue and to enhance the G-M contact bonding [42]. Figure 2(a) shows a scanning electron micrograph (SEM) of a complete TLM structure with a channel width of $W_{CH}$ = 20 μm and $L_{CH}$ ranging from 5 μm till 50 μm. Figure 2 (b) shows a Raman area map of the 2D band intensity in one of the TLM channels, indicated by the blue square in the optical micrograph. A mostly uniform intensity distribution was observed which pointed towards homogenous graphene layers. Figure 2 (c) shows the Raman spectrum of the graphene taken before and after the final device fabrication on SiO$_2$/Si substrate. The red curve is the Raman spectrum of graphene just after the transfer with a ratio of $I_{2D}/I_G$ close to 2, confirming the high quality of the monolayer graphene. The potential presence of some graphene adlayers and grain boundaries result in a small D peak. The TLM structures were vacuum annealed after final device fabrication to enhance the G-M bonding [42]. The black curve in Fig. 2 (c) shows the Raman spectrum of graphene after the annealing process. The 2D/G peak intensity ratio was less than 1 with sharper 2D and G peaks. Such features in Raman spectra of annealed graphene samples point towards strong p-doping of substrate-supported graphene in ambient conditions after the annealing process [43]–[45]. It is worth noticing that the concept of "doping" is here used to describe a change in the Fermi level / carrier density in the graphene due to mirror charges and not the replacement of graphene atoms by dopants as in conventional semiconductor technology. A broad peak observed near the D peak can be attributed to amorphous carbon byproducts formed by carbonization of residual polymer layers on graphene surface [45].

Electrical characterization was carried out in a Lakeshore probe station with a Keithley SCS4200 parameter analyzer in ambient air and in vacuum ($\sim 10^{-6}$ mbar). The samples were kept in Lakeshore for a period of 48 hours to achieve optimum vacuum before measurement. Electrical characterization in ambient atmosphere is carried out by keeping chamber open (humidity $\sim 21\%$, temperature $\sim 300$ K). Figure 3 (a) shows the transfer characteristics [source-drain current ($I_{DS}$) vs back-gate voltage ($V_{BG}$)] of Ni contacted graphene (G – Ni) measured in ambient air. The Dirac voltages ($V_{DIRAC}$) are located at positive $V_{BG}$, indicating a p-doping of the graphene channel [31], [32], [46]. This is in agreement with Raman measurements [Fig. 2(c)], where signatures of p-doping of graphene were observed. This observation can be ascribed to adsorbed water molecules on top of the graphene surface [15]. Graphene acts as an electron donor when water molecules come in contact, leading to the shift of the $V_{DIRAC}$ towards positive gate-voltages, with a charge transfer to graphene per water molecule of approximately 0.002e [47]. We noted a further $V_{DIRAC}$ shift towards positive $V_{BG}$ when $L_{CH}$ increases as summarized in Fig. 3(d). A similar $V_{DIRAC}$ shift was observed by Han et al. in a mechanically exfoliated graphene device and they attributed it to the short channel effects in graphene [48]. However, in our case, short channel effects do not apply. Instead, the effect may be attributed to the polycrystalline nature of the CVD monolayer graphene. Grain boundaries in CVD graphene have been shown to act as active sites for adsorbates like water molecules [49], which can be revealed using HF vapor etching [50]. In the present case, the number of grain boundaries in a device channel should increase with increasing $L_{CH}$, because the device dimensions of the smallest devices are of the same order of magnitude as the grains. Hence, an increased number of grain boundaries in larger devices increases the effect of water adsorbates with $L_{CH}$. Figure 3 (b) is the schematic representation of a TLM structure with varied $L_{CH}$ showing grain boundaries in the CVD graphene.

In vacuum, $V_{DIRAC}$ is significantly shifted compared to ambient air and is observed near -5V, i.e. the graphene channel is slightly n-doped [Fig. 3(c)]. Similar behavior has been observed in Di Bartolomeo et al. [51]. This indicates successful removal of adsorbates by vacuum. This is confirmed by cycling the devices from ambient conditions to vacuum, which shifts $V_{DIRAC}$ consistently back and forth between negative and positive voltages (see section A in the supplementary information). We further found that the $V_{DIRAC}$ difference between different $L_{CH}$ was negligible under vacuum conditions as reported in Fig. 3(d). This supports our assumption concerning the significant role played by grain boundaries and other defects as adsorption sites in ambient condition. These results clearly show that vacuum measurements are required for reliable extraction of $R_{CF}$. Alternatively, controlled encapsulation with dielectrics may be feasible [52], although this is still a field of further research [21].

Having established the large impact of the environment conditions on the device characteristics ($I_{DS}$ vs. $V_{BG}$), we now concentrate on the dependable extraction of the parameters related to G – M contacts. The extraction of $R_{CF}$ for G - Ni contacts [Fig. 3 (a)] when the measurement ($I_{DS}$ – $V_{BG}$) were carried out in ambient air results in erroneous values for varying $V_{BG}$. For example, $R_{CF}$ values of 100 ± 311 Ωμm at $V_{BG}$ = -20 V [Fig. 4(a)] and of -2585 ± 1560 Ωμm at $V_{BG}$ = 20 V [Fig. 4(b)] are extracted, despite achieving high quality fitting (0.97<$R^2$ < 0.999) of the $R_T$ vs $L_{CH}$ curve. Since negative $R_{CF}$ values are not possible in any junction, the source for these errors must be in the extraction technique. In contrast, the TLM measurements under vacuum led to reasonable values of $R_{CF}$; in particular 876 ± 367 Ωμm is extracted at $V_{BG}$ = -20 V [Fig. 4(c)] and 1140 ± 234 Ωμm is extracted at $V_{BG}$ = 20 V [Fig. 4(d)], respectively. This large difference between data in ambient air and high vacuum is also due to the large distortion induced by the air to the $I_{DS}$ - $V_{BG}$ curves in Fig. 4(e) ($L_{CH}$ = 5 μm and $W_{CH}$ = 20 μm). In particular, there is a decrease in the conductivity when the devices are measured in high vacuum, which can be attributed to the removal of water adsorbents from the graphene surface, that provide p-doping to graphene and thus a higher charge density $n_0$ and a higher conductivity. Figure 4 (f) shows the $V_{BG}$ dependence of $R_{CF}$ for the TLM devices measured in high vacuum. $R_{CF}$ clearly is $V_{BG}$ dependent, with a peak at $V_{DIRAC}$, while it decreases as the graphene channel is electrostatically doped by $V_{BG}$. However, despite measuring the devices in vacuum, the extraction of $R_{CF}$ leads to an unusually and, more important, unphysically low value at $V_{BG}$ = -10 V. Current measurements in high vacuum demonstrate a minimum conductance around $V_{BG}$ ~ -5V, with this value varying by less than 0.5 V across the TLM array. To avoid this effect of marginal $V_{DIRAC}$ shifts with varied $L_{CH}$, the minimum conductivity points of all the channels was normalized to 0 V to ensure the same $n_0$ in the graphene channels when comparing the $R_T$ values as discussed in [53]. This normalization process was also applied to the measurements done in air, but the obtained results were unphysical (see section D of the supplementary information).

Figure 5 (a) shows the normalized transfer characteristics [$I_{DS}$ vs. ($V_{BG}$ -$V_{DIRAC}$)] for different $L_{CH}$ of the G - Ni samples measured in vacuum. Figure 5 (b) shows the $R_{CF}$ values extracted through the TLM method after the normalization. The $R_{CF}$ values peak at ($V_{BG}$ -$V_{DIRAC}$)=0, so at the minimum conduction point for graphene. This suggests that the conventional TLM extrapolation technique for $R_{CF}$ is valid only when maintaining the same $n_0$ in graphene channels with different $L_{CH}$. A slight difference in the $n_0$ among the channels leads to large errors in the $R_{CF}$ values [see Figs. 4 (a) and (b)] [48].

Figure 5 (c) shows $R_{CE}$ as a function of $L_{CH}$ at different $V_{BG}$ extracted by using the measurement setup shown in Fig. 1 (a). $R_{CE}$ ranges from 10.5 Ω to 12 Ω with respect to $L_{CH}$ at a fixed $V_{BG}$ demonstrating that $R_{CE}$ is more or less unaffected by the channel resistance outside the contact [33], [54]. The $R_{CE}$ measurement was also carried out in ambient air at $V_{BG}$ =0V [Fig. 5 (d)] and found to be very consistent with the results in vacuum, which indicates that $R_{CE}$ is unaffected by the channel conductivity and, thus, by the measurements conditions (in this case graphene is protected from water molecules by the contact on the top). Finally, $R_{CE}$ is rather independent of $V_{BG}$ and this is most likely because its value is linked to the voltage drop at the end of the contact, where the current density is null and so the $V_{BG}$ dependence this latter is not effective.

$R_{CE}$ and $R_{CF}$ values were then used to calculate $\rho_c$, $L_{TK}$ and $R_{SK}$ (see section C of the supplementary information). Figure 5 (e) shows $L_{TK}$ extracted from measurements done in vacuum as a function of ($V_{BG}$ -$V_{DIRAC}$). $L_{TK}$ is around 1.4 μm at $n_0$= 8.69 x $10^{12}$ cm$^{-2}$ and decreases to 0.9 μm at $n_0$ =1.01 x $10^{12}$ cm$^{-2}$. Figure 5 (f) instead reports the $\rho_c$ as a function of ($V_{BG}$ -$V_{DIRAC}$) and $\rho_c$ is 2.44 x $10^{-5}$ Ωcm$^2$ at $n_0$ = 8.69 x $10^{12}$ cm$^{-2}$ and 7.38 x $10^{-5}$ Ωcm$^2$ at $n_0$ = 1.01 x $10^{12}$ cm$^{-2}$. Error bars indicate upper limits and lower limits in fitting the measured data and they are quite large near $V_{DIRAC}$. Finally, $R_{SH}$ and $R_{SK}$ were extracted to determine the impact of the metal on the properties of graphene under the contact. Figure 5 (g) shows $R_{SK}$ and $R_{SH}$ extracted as a function of $V_{BG}$. $R_{SK}$ for Ni contacts is larger than the $R_{SH}$. Although it has been shown that the p-orbitals of graphene hybridize strongly with Ni d-states [55], the extracted $\rho_c$ and $R_{SK}$ are quite high. This can likely be attributed to nickel-carbide formation at the interface, which can be detrimental for the charge carriers transport through the G-M junction [55]. This is even more important at the Dirac point, where $R_{SK}$ shows a large peak. This is most likely due to the fact that hybridization of graphene orbitals induces small band-gaps at the K-point in the graphene bands [56]. This reduces the graphene charge at the Dirac point, largely impacting the $R_{SK}$ value.

As discussed above, $R_{CF}$ is intrinsically dependent on the charge density in graphene and in the metal. In the case of metal, it is extremely high and in the order of $\sim 10^{21}$ cm$^{-3}$, so it does not limit the current and its effect can be neglected. In the case of graphene, $n_0$ typically varies between $10^{11}$ and $10^{13}$ cm$^{-2}$, therefore $R_{CF}$ is intrinsically dependent on the $R_{SK}$ of graphene under the metal contact [57]. Therefore, it is important to fabricate high quality graphene to lower $R_{SK}$ and thus to reduce $R_{CF}$ [58]. Also selecting the metal materials that increase $n_0$ in the underneath graphene would improve $R_{SK}$ and hence $R_{CF}$.

The methodology described for G-Ni contacts was also applied to the case of graphene – copper (G – Cu) and graphene – gold (G – Au) contacts (see section E in the supplementary information). Furthermore, extensive measurements and extractions were also carried out for TLM devices with G – Au contacts measured in ambient air (see section F in the supplementary information). The measurements in ambient air and vacuum confirm the previous findings, with a large positive $V_{DIRAC}$ under ambient air and smaller $V_{DIRAC}$ values under vacuum (with similar values of approximately 0.7 V for the different $L_{CH}$). The details of these measurements are reported in the section E of the supporting information. The complete set of experimental results is summarized in Tab. 1.

## 4. CONCLUSIONS

The effects of measurement conditions on the extraction of contact related parameters in G-M junctions were investigated in detail using back-gated TLM structures for different metals.

Measurements carried out in ambient conditions, irrespective of the used contact metal, resulted in highly asymmetric transfer curves with positive $V_{DIRAC}$ values, indicating strong p-doping of the graphene channel. Also, a $V_{DIRAC}$ shift in devices with different $L_{CH}$ was observed which is explained on the basis of polycrystalline nature of CVD graphene with non-uniform grain boundaries density.

Vacuum measurements, in contrast, yielded highly symmetric transfer curves for each used metal, which can be reliably used to extract the G-M junction parameters, eliminating one of the main reasons for the scattered values of $R_{CF}$ (and $\rho_c$) reported in the literature.

Discrepancies related to the extraction of $R_{CF}$ (and $\rho_c$) via the TLM method were also discussed rigorously. In particular, to extract dependable $R_{CF}$ values, the $n_0$ in the graphene between the contacts should be kept constant for the different devices ($L_{CH}$), and hence small differences in $V_{DIRAC}$ position should be compensated. $R_{CF}$ is strongly dependent on the $n_0$ in the graphene underneath the metal, with the lowest value achieved for gold contacts. The present study highlights the importance of a careful extraction of the contact related parameter in G-M junctions.

## Acknowledgements

Financial support from the European Commission (Graphene Flagship, 785219, 881603), the German Ministry for Education and Research, BMBF (GIMMIK, 03XP0210F) and the Italian Minister of Education, University and Research, MIUR (Five2D, 2017SRYEJH) is gratefully acknowledged.

## References

[1]     K. S. Novoselov *et al.*, "Electric field effect in atomically thin carbon films," *Science*, vol. 306, no. 5696, pp. 666–669, 2004.

[2]     Y.-M. Lin, K. A. Jenkins, A. Valdes-Garcia, J. P. Small, D. B. Farmer, and P. Avouris, "Operation of graphene transistors at gigahertz frequencies," *Nano Lett.*, vol. 9, no. 1, pp. 422–426, 2008.

[3]     Y.-M. Lin *et al.*, "100-GHz transistors from wafer-scale epitaxial graphene," *Science*, vol. 327, no. 5966, pp. 662–662, 2010.

[4]     L. Liao *et al.*, "High-speed graphene transistors with a self-aligned nanowire gate," *Nature*, vol. 467, no. 7313, pp. 305–308, 2010.

[5]     G. Fiori and G. Iannaccone, "Insights on radio frequency bilayer graphene FETs," in *Electron Devices Meeting (IEDM), 2012 IEEE International*, 2012, pp. 17–3, Accessed: Sep. 17, 2017. [Online]. Available: http://ieeexplore.ieee.org/abstract/document/6479059/.

[6]     H. Pandey, S. Kataria, A. Gahoi, and M. C. Lemme, "High Voltage Gain Inverters From Artificially Stacked Bilayer CVD Graphene FETs," *IEEE Electron Device Lett.*, vol. 38, no. 12, pp. 1747–1750, 2017.

[7]     S. Vaziri *et al.*, "Going ballistic: Graphene hot electron transistors," *Solid State Commun.*, vol. 224, pp. 64–75, 2015.

[8]     T. Mueller, F. Xia, and P. Avouris, "Graphene photodetectors for high-speed optical communications," *Nat Photon*, vol. 4, pp. 297–301, 2010.

[9]     M. C. Lemme *et al.*, "Gate-Activated Photoresponse in a Graphene p–n Junction," *Nano Lett.*, vol. 11, no. 10, pp. 4134–4137, Oct. 2011, doi: 10.1021/nl2019068.

[10]    F. Bonaccorso, Z. Sun, T. Hasan, and A. C. Ferrari, "Graphene photonics and optoelectronics," *Nat. Photonics*, vol. 4, no. 9, pp. 611–622, 2010.

[11]    M. Furchi *et al.*, "Microcavity-integrated graphene photodetector," *Nano Lett.*, vol. 12, no. 6, pp. 2773–2777, 2012.

[12]    S. Riazimehr *et al.*, "High Responsivity and Quantum Efficiency of Graphene/Silicon Photodiodes Achieved by Interdigitating Schottky and Gated Regions," *ACS Photonics*, vol. 6, no. 1, pp. 107–115, Jan. 2019, doi: 10.1021/acsphotonics.8b00951.

[13]    A. D. Smith *et al.*, "Electromechanical piezoresistive sensing in suspended graphene

membranes," *Nano Lett.*, vol. 13, no. 7, pp. 3237–3242, 2013.

[14]  C. Chen and J. Hone, "Graphene nanoelectromechanical systems," *Proc. IEEE*, vol. 101, no. 7, pp. 1766–1779, 2013.

[15]  A. D. Smith *et al.*, "Resistive graphene humidity sensors with rapid and direct electrical readout," *Nanoscale*, vol. 7, no. 45, pp. 19099–19109, 2015.

[16]  A. D. Smith *et al.*, "Piezoresistive Properties of Suspended Graphene Membranes under Uniaxial and Biaxial Strain in Nanoelectromechanical Pressure Sensors," *ACS Nano*, vol. 10, no. 11, pp. 9879–9886, Nov. 2016, doi: 10.1021/acsnano.6b02533.

[17]  X. Fan *et al.*, "Graphene ribbons with suspended masses as transducers in ultra-small nanoelectromechanical accelerometers," *Nat. Electron.*, vol. 2, pp. 394–404, Sep. 2019, doi: 10.1038/s41928-019-0287-1.

[18]  O. Graydon, "Graphene: Terahertz modulator," *Nat. Photonics*, vol. 9, no. 12, pp. 780–780, 2015.

[19]  B. Sensale-Rodriguez *et al.*, "Broadband graphene terahertz modulators enabled by intraband transitions," *Nat. Commun.*, vol. 3, p. 780, 2012.

[20]  Q.-Y. Wen *et al.*, "Graphene based all-optical spatial terahertz modulator," *Sci. Rep.*, vol. 4, 2014.

[21]  D. Neumaier, S. Pindl, and M. C. Lemme, "Integrating graphene into semiconductor fabrication lines," *Nat. Mater.*, vol. 18, no. 6, p. 525, Jun. 2019, doi: 10.1038/s41563-019-0359-7.

[22]  K. Nagashio and A. Toriumi, "Density-of-states limited contact resistance in graphene field-effect transistors," *Jpn. J. Appl. Phys.*, vol. 50, no. 7, p. 0108, 2011.

[23]  K. Nagashio, T. Nishimura, K. Kita, and A. Toriumi, "Metal/graphene contact as a performance killer of ultra-high mobility graphene analysis of intrinsic mobility and contact resistance," in *Electron Devices Meeting (IEDM), 2009 IEEE International*, 2009, pp. 1–4, Accessed: Dec. 30, 2013. [Online]. Available: http://ieeexplore.ieee.org/xpls/abs_all.jsp?arnumber=5424297.

[24]  J. T. Smith, A. D. Franklin, D. B. Farmer, and C. D. Dimitrakopoulos, "Reducing Contact Resistance in Graphene Devices through Contact Area Patterning," *ACS Nano*, vol. 7, no. 4, pp. 3661–3667, Apr. 2013, doi: 10.1021/nn400671z.

[25]  L. Anzi *et al.*, "Ultra-low contact resistance in graphene devices at the Dirac point," *2D Mater.*, vol. 5, no. 2, p. 025014, 2018, doi: 10.1088/2053-1583/aaab96.

[26]  V. Passi *et al.*, "Ultralow Specific Contact Resistivity in Metal–Graphene Junctions via Contact Engineering," *Adv. Mater. Interfaces*, vol. 6, no. 1, p. 1801285, 2019, doi: 10.1002/admi.201801285.

[27]  F. Xia, V. Perebeinos, Y. Lin, Y. Wu, and P. Avouris, "The origins and limits of metal-graphene junction resistance," *Nat. Nanotechnol.*, vol. 6, no. 3, pp. 179–184, 2011.

[28]  J. S. Moon *et al.*, "Ultra-low resistance ohmic contacts in graphene field effect transistors," *Appl. Phys. Lett.*, vol. 100, no. 20, pp. 203512–203512, 2012.

[29]  A. Gahoi, S. Wagner, A. Bablich, S. Kataria, V. Passi, and M. C. Lemme, "Contact resistance study of various metal electrodes with CVD graphene," *Solid-State Electron.*, vol. 125, pp. 234–239, 2016.

[30] A. Meersha et al., "Record low metal—(CVD) graphene contact resistance using atomic orbital overlap engineering," in *Electron Devices Meeting (IEDM), 2016 IEEE International*, 2016, pp. 5–3, Accessed: Sep. 08, 2017. [Online]. Available: http://ieeexplore.ieee.org/abstract/document/7838352/.

[31] S. Wang et al., "A more reliable measurement method for metal/graphene contact resistance," *Nanotechnology*, vol. 26, no. 40, p. 405706, 2015.

[32] S. Wang et al., "Characterization of the quality of metal–graphene contact with contact end resistance measurement," *Appl. Phys. A*, vol. 122, no. 7, pp. 1–7, 2016.

[33] S. Venica, F. Driussi, A. Gahoi, P. Palestri, M. C. Lemme, and L. Selmi, "On the Adequacy of the Transmission Line Model to Describe the Graphene–Metal Contact Resistance," *IEEE Trans. Electron Devices*, vol. 65, no. 4, pp. 1589–1596, 2018.

[34] M. König et al., "Accurate graphene-metal junction characterization," *IEEE J. Electron Devices Soc.*, vol. 7, pp. 219–226, 2019.

[35] F. Urban, G. Lupina, A. Grillo, N. Martucciello, and A. Di Bartolomeo, "Contact resistance and mobility in back-gate graphene transistors," *Nano Express*, vol. 1, no. 1, p. 010001, 2020.

[36] F. Driussi et al., "Improved understanding of metal–graphene contacts," *Microelectron. Eng.*, p. 111035, 2019.

[37] A. Di Bartolomeo et al., "Charge transfer and partial pinning at the contacts as the origin of a double dip in the transfer characteristics of graphene-based field-effect transistors," *Nanotechnology*, vol. 22, no. 27, p. 275702, 2011.

[38] G. K. Reeves and H. B. Harrison, "Obtaining the specific contact resistance from transmission line model measurements," *IEEE Electron Device Lett.*, vol. 3, no. 5, pp. 111–113, 1982.

[39] G. Giovannetti, P. A. Khomyakov, G. Brocks, V. M. Karpan, J. Van den Brink, and P. J. Kelly, "Doping graphene with metal contacts," *Phys. Rev. Lett.*, vol. 101, no. 2, p. 026803, 2008.

[40] S. Kataria et al., "Chemical vapor deposited graphene: From synthesis to applications," *Phys. Status Solidi A*, vol. 211, no. 11, pp. 2439–2449, 2014.

[41] Y. Wang et al., "Electrochemical delamination of CVD-grown graphene film: toward the recyclable use of copper catalyst," *ACS Nano*, vol. 5, no. 12, pp. 9927–9933, 2011.

[42] W. S. Leong, C. T. Nai, and J. T. Thong, "What Does Annealing Do to Metal–Graphene Contacts?," *Nano Lett.*, vol. 14, no. 7, pp. 3840–3847, 2014.

[43] S. Wagner et al., "Noninvasive scanning Raman spectroscopy and tomography for graphene membrane characterization," *Nano Lett.*, vol. 17, no. 3, pp. 1504–1511, 2017.

[44] Z. H. Ni et al., "The effect of vacuum annealing on graphene," *J. Raman Spectrosc. Int. J. Orig. Work Asp. Raman Spectrosc. High. Order Process. Also Brillouin Rayleigh Scatt.*, vol. 41, no. 5, pp. 479–483, 2010.

[45] J. Hong, M. K. Park, E. J. Lee, D. Lee, D. S. Hwang, and S. Ryu, "Origin of new broad Raman D and G peaks in annealed graphene," *Sci. Rep.*, vol. 3, p. 2700, 2013.

[46] L. Anzi et al., "Ultra-low contact resistance in graphene devices at the Dirac point," *2D Mater.*, vol. 5, no. 2, p. 025014, 2018.

[47]  O. Leenaerts, B. Partoens, and F. M. Peeters, "Water on graphene: Hydrophobicity and dipole moment using density functional theory," *Phys. Rev. B*, vol. 79, no. 23, p. 235440, 2009.

[48]  S.-J. Han, Y. Sun, A. A. Bol, W. Haensch, and Z. Chen, "Study of channel length scaling in large-scale graphene FETs," in *2010 Symposium on VLSI Technology*, 2010, pp. 231–232.

[49]  A. W. Cummings *et al.*, "Charge Transport in Polycrystalline Graphene: Challenges and Opportunities," *Adv. Mater.*, vol. 26, no. 30, pp. 5079–5094, 2014, doi: 10.1002/adma.201401389.

[50]  X. Fan *et al.*, "Direct observation of grain boundaries in graphene through vapor hydrofluoric acid (VHF) exposure," *Sci. Adv.*, vol. 4, no. 5, p. eaar5170, 2018.

[51]  A. Di Bartolomeo *et al.*, "Graphene field effect transistors with niobium contacts and asymmetric transfer characteristics," *Nanotechnology*, vol. 26, no. 47, p. 475202, 2015.

[52]  A. A. Sagade *et al.*, "Highly air stable passivation of graphene based field effect devices," *Nanoscale*, vol. 7, no. 8, pp. 3558–3564, 2015, doi: 10.1039/C4NR07457B.

[53]  S. Venica *et al.*, "Reliability analysis of the metal-graphene contact resistance extracted by the transfer length method," in *2018 IEEE International Conference on Microelectronic Test Structures (ICMTS)*, 2018, pp. 57–62.

[54]  S. S. Cohen and G. S. Gildenblat, *Metal–Semiconductor Contacts and Devices*, vol. 13. Academic Press, 2014.

[55]  A. Dahal and M. Batzill, "Graphene–nickel interfaces: a review," *Nanoscale*, vol. 6, no. 5, pp. 2548–2562, 2014.

[56]  P. Khakbaz *et al.*, "DFT study of graphene doping due to metal contacts," in *2019 International Conference on Simulation of Semiconductor Processes and Devices (SISPAD)*, Sep. 2019, pp. 1–4, doi: 10.1109/SISPAD.2019.8870456.

[57]  T. Cusati *et al.*, "Electrical properties of graphene-metal contacts," *Sci. Rep.*, vol. 7, 2017, Accessed: Sep. 17, 2017. [Online]. Available: https://www.ncbi.nlm.nih.gov/pmc/articles/PMC5506027/.

[58]  S. Wittmann *et al.*, "Dielectric surface charge engineering for electrostatic doping of graphene," *Revis. ACS Appl. Electron. Mater.*, 2020.

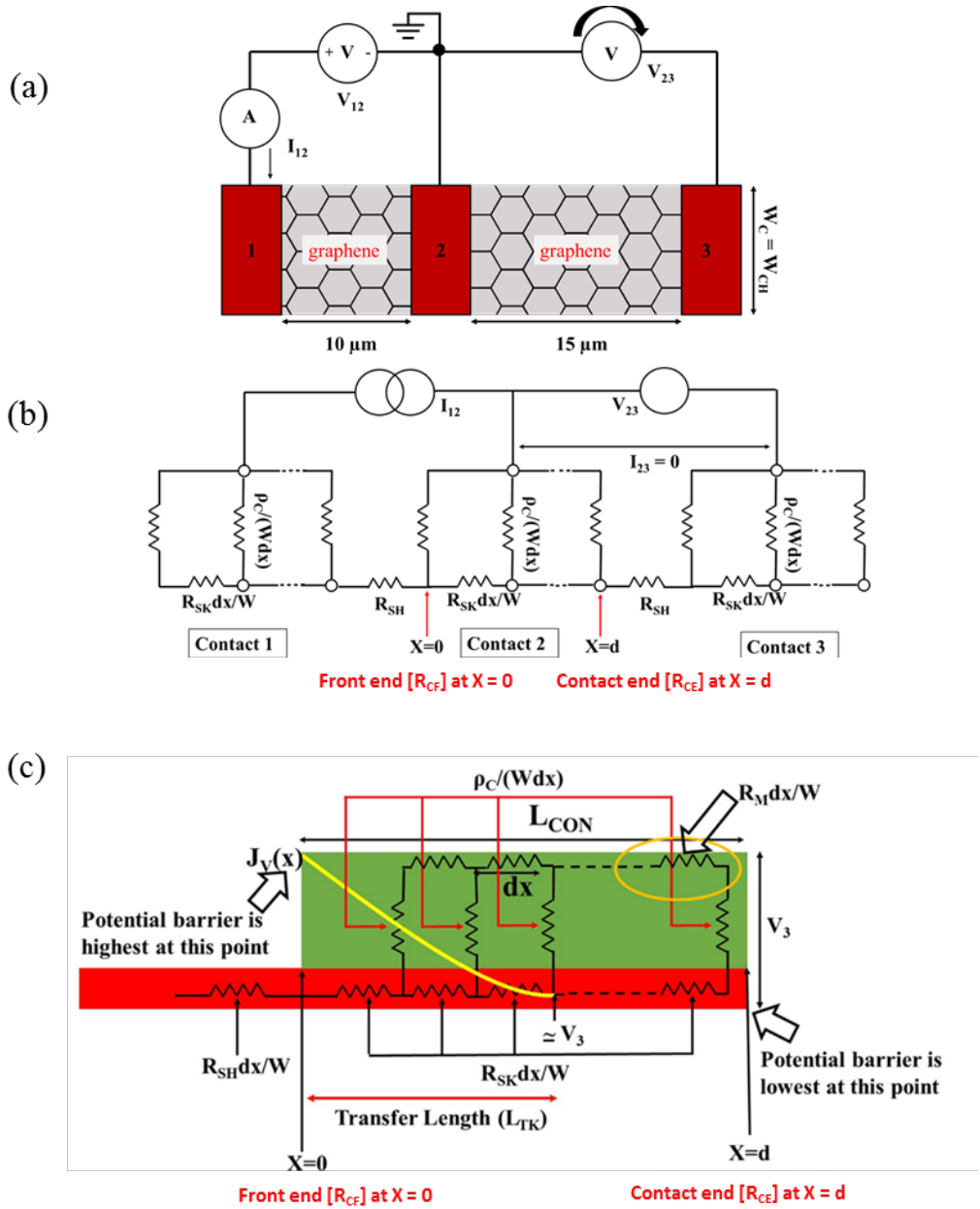

| | (d) Important Notation | |
|---|---|
| $R_T$ (Ω) - total resistance | $R_{SK}$ (Ω/□) - sheet resistance of the graphene under metal contact |
| d (μm) - contact length | $\rho_c$ (Ωcm$^2$) - specific contact resistivity |
| $R_C$ (Ωμm) - contact resistance | $R_{CE}$ (Ω) - contact end resistance |
| $L_{TK}$ (μm) - transfer length | $R_{SH}$ (Ω/□) - sheet resistance of the graphene between the contacts |
| $W_{CH}$ (μm) - channel width | $R_{CF}$ (Ωμm) - contact front resistance |
| $L_{CH}$ (μm) - channel length | |

**Figure 1. (a) Measurement setup for TLM structures. (b) Equivalent circuit diagram describing the distributed resistive components in a TLM structure. (c) Distributed circuit diagram of the transmission line model to describe the G-M contact. X=0 is front of the contact and X=d is end of the contact (d = length of the contact). The horizontal voltage drop due to the graphene resistance leads to the current crowding at the G-M junction. (d) Important notation used in this paper.**

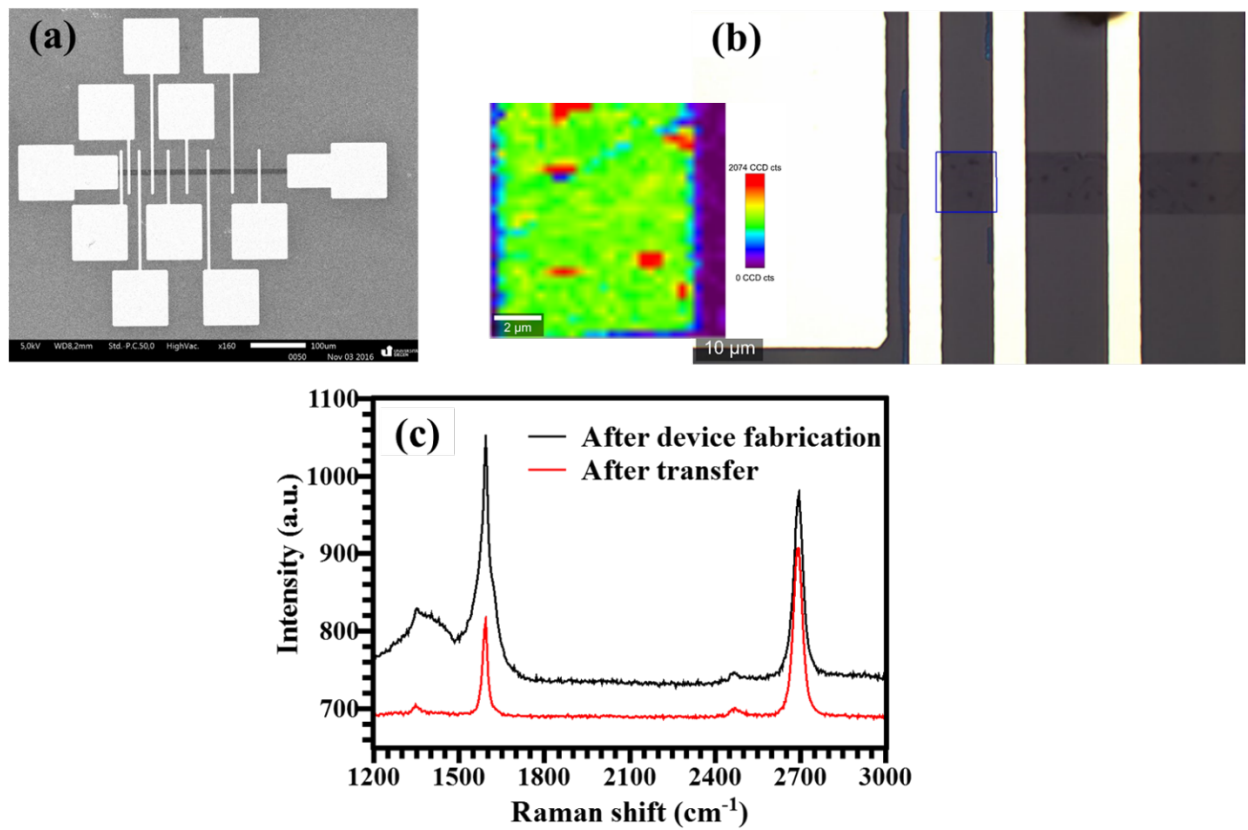

**Figure 2.** (a) Scanning electron micrograph (SEM) of a graphene FET array (TLM structures. (b) Raman area map of 2D peak in the area depicted by the blue square in the optical micrograph of the device. (c) Raman spectrum of a CVD graphene transferred on a SiO$_2$/Si substrate.

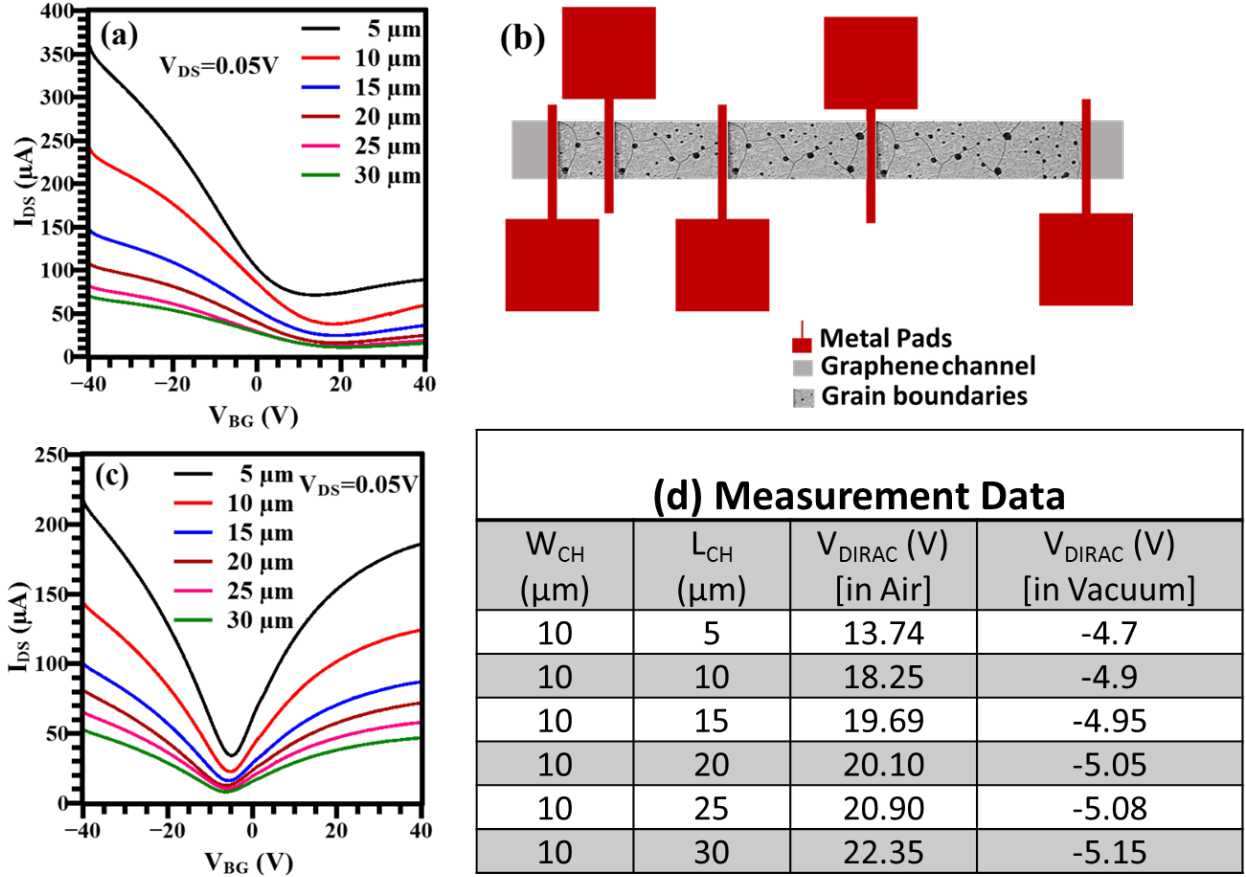

**Figure 3. (a)** Transfer characteristics ($I_{DS}$-$V_{BG}$) measured in ambient atmosphere (humidity ~21%, temperature ~ 300K). **(b)** Schematic structure of a TLM structure depicting grain boundaries or line defects in the CVD graphene. **(c)** Transfer characteristics ($I_{DS}$-$V_{BG}$) measured in high vacuum (humidity~0%, temperature ~ 300 K, Pressure ~ $10^{-7}$ mbar). **(d)** Summary of a data in a tabular form focusing on $V_{DIRAC}$ position measured in ambient air and in vacuum.

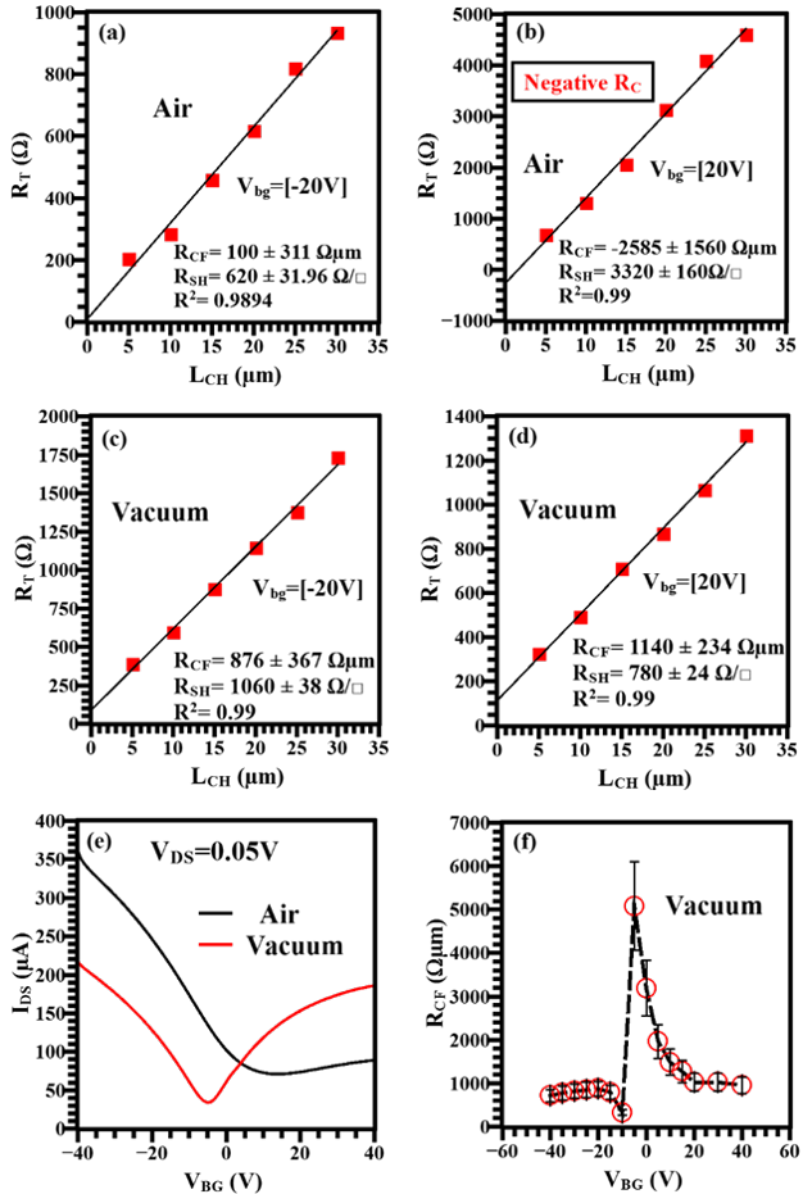

**Figure 4.** Extrapolation of $R_{CF}$ and $R_{SH}$ by using the transfer characteristics curves of graphene - nickel samples [Fig. 3(a)] measured in ambient atmosphere at gate biases of -20 V (a, p-doped) and 20 V (b, n-doped). Red squares represent measured total resistance; black line, is the linear fitting curve. $R_{CF}$ and $R_{SH}$ are also extrapolated by using the transfer characteristics curves [Fig. 3 (c)] measured in high vacuum at gate biases of -20 V (c) and 20 V (d). (e) Comparison of transfer characteristics measured in air and vacuum. (f) $R_{CF}$ as a function of a $V_{BG}$.

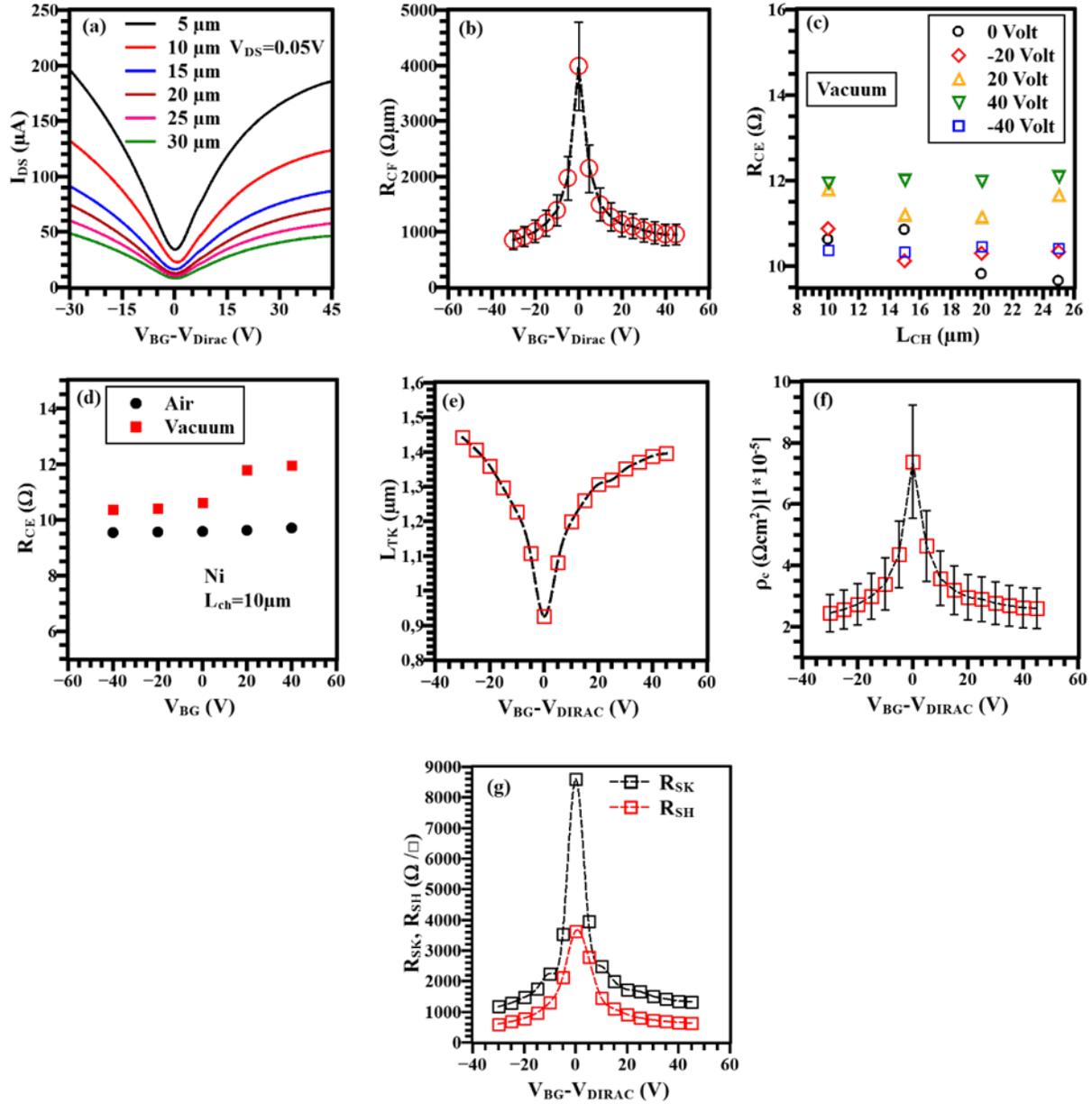

**Figure 5. (a) Transfer characteristics of G-Ni TLM structures after normalizing to $V_{DIRAC}=0$ V the curves of Fig. 3 (c). (b) $R_{CF}$ as a function of a $V_{BG} - V_{DIRAC}$. (c) $R_{CE}$ as a function of $L_{CH}$ measured at different $V_{BG}$. (d) $R_{CE}$ as a function of $V_{BG}$ in ambient atmosphere and in vacuum. (e) $L_{TK}$ and (f) $\rho_c$ as a function of $V_{BG} - V_{DIRAC}$. (g) $R_{SK}$ and $R_{SH}$ as a function of $V_{BG} - V_{DIRAC}$.**

**Table 1: - Summary of results**

| Graphene – Metal Contact | ($R_{CF}$) (Ωμm) | $L_{TK}$) (μm) | $R_{SK}$ (Ω/□) | $R_{SH}$ (Ω/□) | $\rho_c$ (Ωcm²) |
|---|---|---|---|---|---|
| **G-Ni ( n = 8.69 x 10$^{12}$ cm$^{-2}$ )** | 853 ± 171 | 1.45 | 1174 | 608 | 2.44 x 10$^{-5}$ |
| G-Cu ( n = 1.15 x 10$^{13}$ cm$^{-2}$) | 860 ± 172 | 1.63 | 417 | 729 | 1.1 x 10$^{-5}$ |
| G-Au ( n = 1.07 x 10$^{13}$ cm$^{-2}$) | 395 ± 79 | 1.68 | 376 | 425 | 7.16 x 10$^{-6}$ |
| G-Ni ( n = 1.01 x 10$^{12}$ cm$^{-2}$ ) | 3984 ± 799 | 0.93 | 8599 | 3642 | 7.38 x 10$^{-5}$ |
| G-Cu ( n = 1.34 x 10$^{12}$ cm$^{-2}$) | 3664 ± 733 | 0.97 | 3359 | 3540 | 3.5 x 10$^{-5}$ |
| G-Au ( n = 1.06 x 10$^{12}$ cm$^{-2}$) | 1303 ± 261 | 0.97 | 1081 | 2700 | 1.56 x 10$^{-5}$ |